\documentclass[prd,twocolumn,showpacs,floatfix,amsmath,amssymb,floatfix]{revtex4}
\usepackage{graphicx,color,dcolumn,booktabs,bm,mathrsfs}
\usepackage{longtable,lscape}
\usepackage{txfonts}
\usepackage{overpic}
\usepackage{amssymb}
\usepackage{indentfirst}
\usepackage{feynmf}   
\usepackage{slashed}  
\usepackage{cases}
\usepackage{color}
\usepackage{multirow}
\usepackage{epstopdf}

\graphicspath{{Figures/}} %
\usepackage[colorlinks,
                      citecolor=blue,
                      anchorcolor=red,
                      menucolor=red,
                      linkcolor=red,
                      filecolor=red,
                      runcolor=red,
                      urlcolor=blue,
                      frenchlinks=red]{hyperref}

\usepackage[]{changes}
	\definechangesauthor[name={JH}, color=orange]{JH}
	\setremarkmarkup{(#2)}
\begin{document}

\title{Production of $P-$wave charmed and charmed-strange mesons in pion and kaon induced reactions}

\author{Jing Liu$^1$}\email{ jingliu@seu.edu.cn}
\author{Qi Wu$^1$}\email{ wuq@seu.edu.cn}
\author{Jun He $^2$}\email{junhe@njnu.edu.cn}
\author{Dian-Yong Chen $^1$ \footnote{Corresponding author}} \email{chendy@seu.edu.cn}
\author{Takayuki Matsuki $^{3,4}$}\email{matsuki@tokyo-kasei.ac.jp}
\affiliation{
$^1$ School of Physics, Southeast University,  Nanjing 210094, China\\
$^2$Department of Physics and Institute of Theoretical Physics,
Nanjing Normal University, Nanjing 210097, China\\
$^3$Tokyo Kasei University, 1-18-1 Kaga, Itabashi, Tokyo 173-8602, Japan \\
$^4$Theoretical Research Division, Nishina Center, RIKEN, Wako, Saitama 351-0198, Japan
}

\begin{abstract}
In the present work, we investigate the production of $P-$wave charmed or charmed-strange mesons in the pion or kaon induced reactions on a proton target.  The total cross sections as well as the differential cross sections depending on the scattering angle are evaluated in an effective Lagrangian approach. Our estimations indicate that magnitude of the cross sections strongly depends on the model parameter but such dependence can be almost completely canceled for the cross section ratios. These model independent ratios can be taken as a good criterion of the validity of heavy quark limit in the charmed region, which is helpful to understand $P$-wave charmed and charmed-strange mesons.
\end{abstract}

\pacs{13.87.Ce, 13.75.Gx, 13.75.Jz }

\maketitle

\section{Introduction}
\label{sec:introduction}
In the past decades,  a series of charmed and charmed-strange resonances have been observed, which makes the charmed and charmed-strange families abundant.  Most of these resonances can be well categorized into charmed or charmed-strange meson spectrum predicted in the constituent quark model. The meson spectrum was predicted  long time ago, by such as the relativistic quark model proposed by Godfrey and Isgur ~\cite{Godfrey:1985xj}. As for the charmed mesons, the  $S$ wave ground states, $D$ and $D^\ast$, were observed as early as in the 1970s.
In 1985, the ARGUS Collaboration reported a new resonance near 2420 MeV in the $D^\ast \pi$ invariant mass spectrum~\cite{Albrecht:1985as}. Later in the $D \pi$ invariant mass spectrum, another $P$-wave charmed meson, $D_2^\ast (2460)$ was also reported at ARGUS~\cite{Albrecht:1988dj}. Further analysis of the ARGUS data indicates that the broad structure reported in Ref.~\cite{Albrecht:1985as} is actually the superposition of two relatively narrow states, $D_2^\ast(2460)$ and $D_1 (2420)$.  Until 2004, the rest of $P$-wave ground charmed mesons, $D_1(2430)$ and $D_0(2400)$, was reported by the FOCUS~\cite{Link:2003bd} and Belle Collaborations~\cite{Abe:2003zm}, respectively.  Since then, the $P$-wave ground charmed mesons have been established, and the observed masses of these states are consistent with expectations of the relativistic quark model~\cite{Godfrey:1985xj}.

\begin{figure}[t]
\includegraphics[scale=0.65]{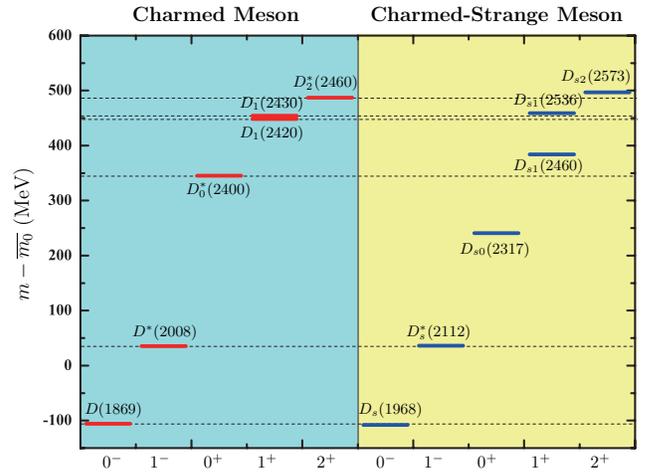}
\caption{A comparison of the residual masses of the lowest $S-$ and $P-$ wave charmed and charmed-strange mesons. The $\overline{m_0}$ is defined as $ (m_D+ 3m_{D^\ast})/4$ and $ (m_{D_s}+ 3m_{D_s^\ast})/4$ for charmed and charmed-strange  mesons, respectively. The  dashed lines indicate the residual masses of the charmed mesons extended to the charmed-strange mesons.\label{Fig:compare}}
\end{figure}

However, establishment of the $P$-wave ground charmed-strange mesons is not as smooth as the charmed mesons. In 1988, the $P$-wave charmed-strange meson, $D_{s1}(2536)$, was observed in the $D_s \gamma$ invariant mass spectrum of a $\bar{\nu} N$ collision process \cite{Asratian:1987rb}. The $2^+ $ charmed-strange meson, named $D_{s2}^{\ast} (2573)$\added{,} was first observed in the $D^0 K^+$ invariant mass spectrum of a $B$ decay process.  In 2003, two charmed strange states, named $D_{s0}^\ast (2317)$ and $D_{s1}^\prime (2460)$, were first observed in the $D_s^+ \pi^0$ invariant mass spectrum of a $B$ decay process by the {\it BABAR} Collaboration~\cite{Aubert:2003fg} and  in the $D_s^{\ast + } \pi^0$ invariant mass spectrum of a $B$ decay process by the CLEO Collaboration~\cite{Besson:2003cp}, respectively.  The $J^P$ quantum numbers of $D_{s0}^\ast(2317)$ and $D_{s1}(2460)$ indicate that they could be good candidates of $P$-wave charmed-strange mesons. However, the observed masses of these two states are far below the quark model expectation~\cite{Godfrey:1985xj}. In order to show the flavor symmetry of up, down and strange quarks, we define a residual mass as the difference of meson mass and $\overline{m_0}$, where $\overline{m_0}= (m_{D}+3 m_{D^\ast})/4$ and $\overline{m_0}= (m_{D_s}+3 m_{D_s^\ast})/4$ for charmed and charmed-strange mesons, respectively.  As shown in Fig.~\ref{Fig:compare}, we present a comparison of the residual masses of the lowest $S-$ and $P-$wave charmed and charmed-strange mesons . As for the $S-$wave states, the residual masses are almost the same for all light flavors, which reflects the flavor symmetry.  As for the $P-$wave states, such a flavor symmetry is kept for $D_1^\prime(2430)/ D_{s1}^\prime (2536)$ and $D_2^{\ast}(2460)/D_{s2}(2573)$. However, if one takes $D_{s0}^\ast (2317)$ and $D_{s1}(2460)$ as the strange partners of  $D_0^\ast(2400)$ and $D_1(2420)$, one can find that the residual masses of $D_{s0}^\ast (2317)$ and $D_{s1}(2460)$ are much smaller than those of  $D_0^\ast(2400)$ and $D_1(2420)$.

The particular properties of $D_{s0}^\ast(2317)$ and $D_{s1}(2460)$ stimulate theorists to great interests. It is interesting to notice that the observed masses of  $D_{s0}^\ast(2317)$ and $D_{s1}(2460)$  are just below the thresholds of $D K$ and $D^\ast K$, respectively. Thus, it was natural to evaluate the possibility of interpreting both states as deuteronlike molecular states from different aspects, such as mass spectrum~\cite{Xie:2010zza, Zhang:2006ix, Bicudo:2004dx}, decay behavior~\cite{Faessler:2007gv, Faessler:2007us, Cleven:2014oka, Xiao:2016hoa}, and production process~\cite{Datta:2003re}. However, if $D_{s0}^\ast(2317)$ and $D_{s1}(2460)$ are considered as a molecular state, one has to answer the question ``where are $P$-wave charmed strange mesons with $J^P=0^+$ and $1^+$?''.  However, there are no evidence of  additional states corresponding to $D_{s0}^\ast$ and $D_{s1}$.  In Refs.~\cite{Lutz:2008zz, Hwang:2004cd, Song:2015nia}, the mass discrepancies were interpreted as the coupled-channel effect and the relativistic quark model with fine tuning parameters could reproduce the masses of $D_{s0}^\ast(2317)$ and $D_{s1}(2460)$~\cite{Liu:2013maa}.   The decay behaviors investigating the charmed-strange meson scenario also indicates that the observed properties of $D_{s0}^\ast (2317)$ and $D_{s1}(2460)$ could be understood~\cite{Liu:2006jx, Lu:2006ry, Wang:2006mf, Fajfer:2015zma, Dai:2003yg, Colangelo:2003vg, Colangelo:2005hv}.

To further test our understanding of the $P$-wave charmed and charmed-strange mesons, their properties should be investigated in more aspects. If one carefully checks the experimental production process of these heavy-light mesons, one can find most of these charmed and charmed-strange mesons are produced via $e^+e^-$ annihilation, photon production, and $B$ meson decay processes.  Actually, the charmed and charmed strange mesons could also be produced in the pion/kaon induced processes, such as a reaction on a nuclear target. Up to date, we have high energy pion and kaon beams with high quality, such as J-PARC~\cite{Nagae:2008zz},  COMPASS~\cite{Nerling:2012er}, OKA@U-70~\cite{Obraztsov:2016lhp} and SPS@CERN~\cite{Velghe:2016jjw}. These facilities make the charmed and charmed meson production feasible in the hadronic production. In the present work, we investigate the $P$-wave charmed and charmed strange mesons by pion and kaon induced reactions on a proton target, which could provide some useful information for further experimental measurements. On the other hand, the  measurements could also provide some critical tests for our understanding of these charmed and charmed-strange mesons, especially for $D_{s0}^{\ast} (2317)$ and $D_{s1}(2460)$.

This work is organized as follows. After the Introduction, we present our investigation of the pion and kaon induced production on a proton target at hadronic level and the related amplitudes are obtained by the effective Lagrangians constructed in the heavy quark limit and chiral symmetry. In Sec.~\ref{Sec:Num}, we present our numerical results and some discussions, and last section is devoted to a short summary.

\section{Pion (Kaon) induced production on a proton target}

\begin{figure}[ht]
\begin{tabular}{ccc}
  \centering
 \includegraphics[width=4.00cm]{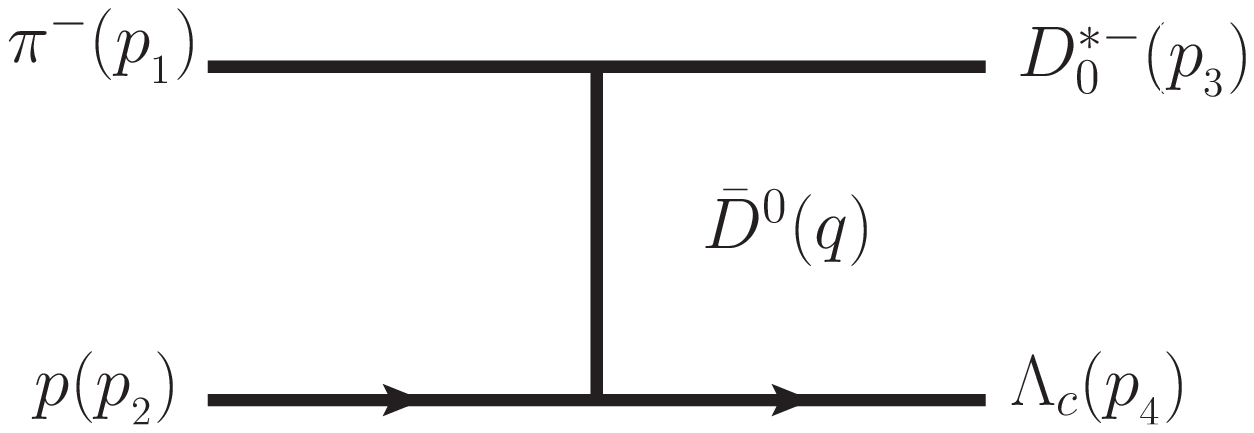}&
  \includegraphics[width=4.00cm]{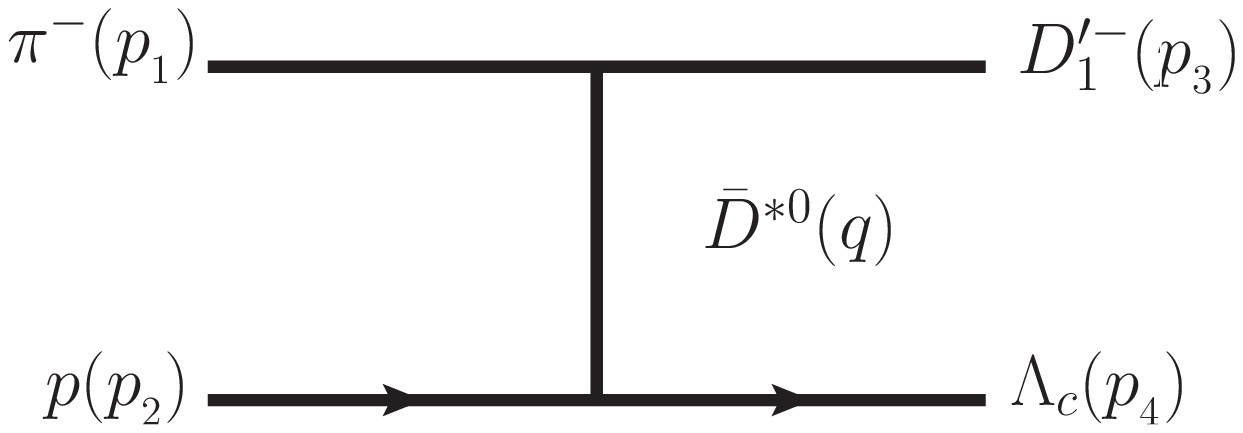}&
 \\
 \\
 $(a)$ & $(b)$\\
 \includegraphics[width=4.00cm]{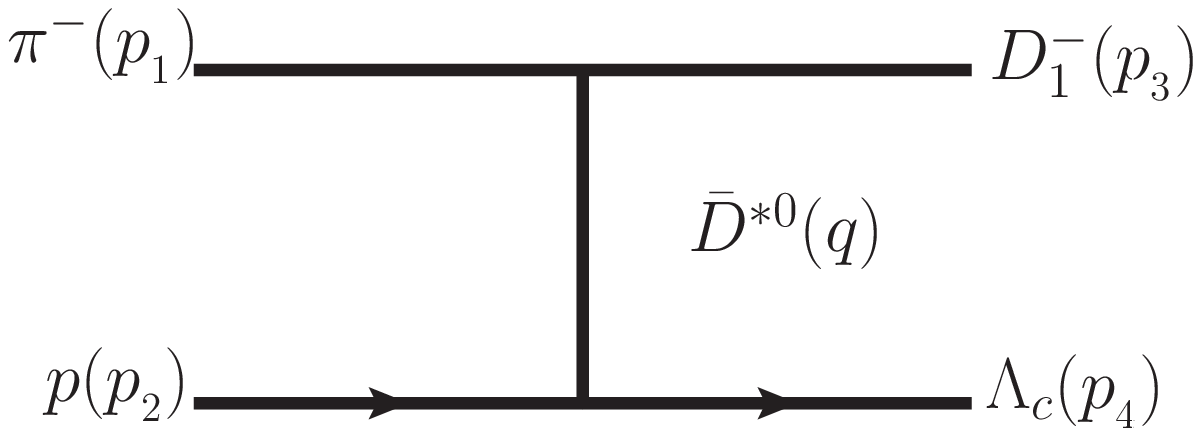}&
 \includegraphics[width=4.00cm]{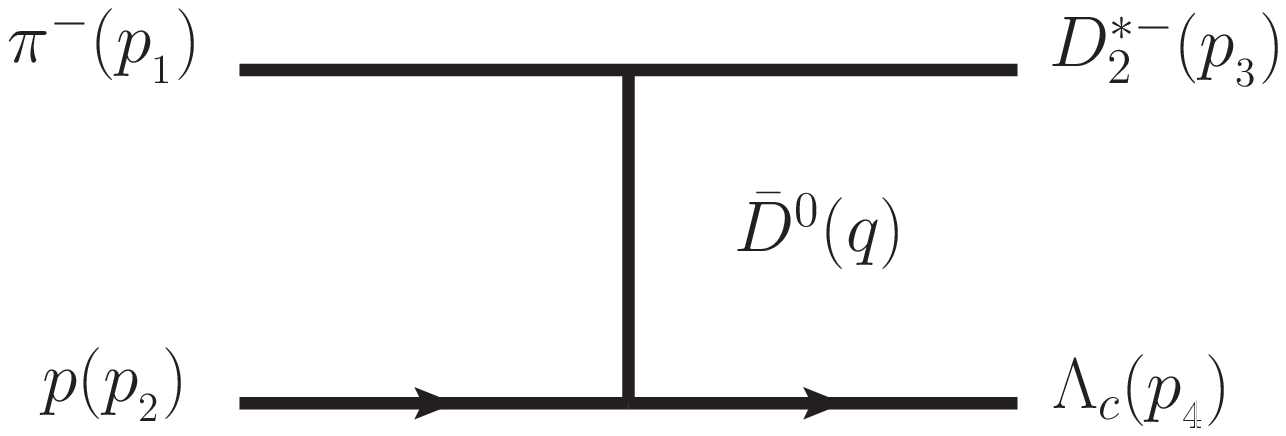}&
 \\
 \\
 $(c)$ & $(d)$&\\
 \includegraphics[width=4.00cm]{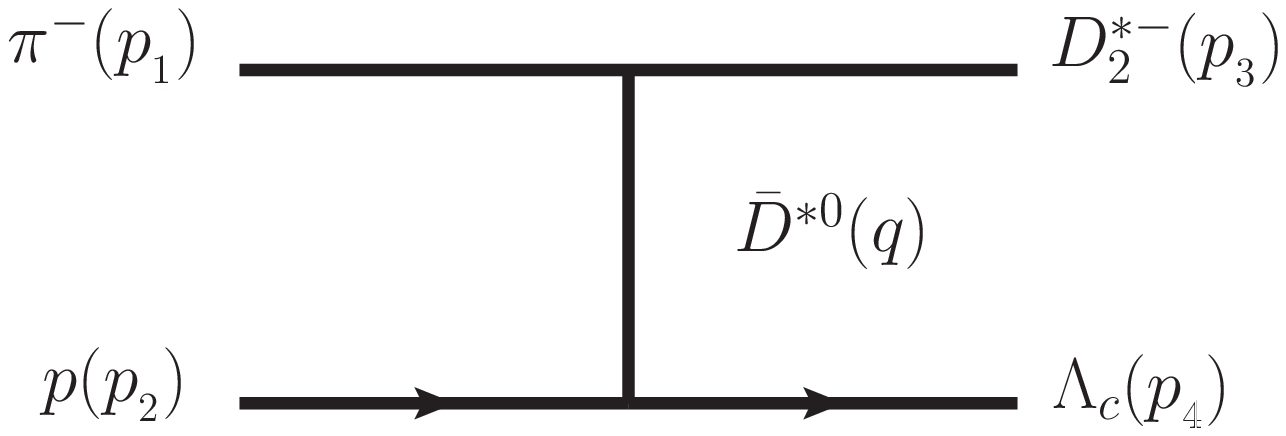}\\
 $(e)$
 \end{tabular}
\caption{Diagrams contributing to $\pi^- p \to D_0^{\ast-} \Lambda_c$ (a), $\pi^- p \to
 D_1^{-}  \Lambda_c$ (b), $\pi^- p \to
 D_1^{\prime -}  \Lambda_c$ (c), and $\pi^- p \to
 D_2^{-}  \Lambda_c$ ((d) and (e)). The diagrams related to the $P$-wave charmed-strange meson production could be obtained by replacing $\pi^-$ with $K^-$.  \label{Fig:Feyn}}
\end{figure}

In the present work, we estimate the $P$-wave production process at hadronic level and an effective Lagrangian approach is adopted to depict the hadron interactions. Here, the effective Lagrangians are constructed in the heavy quark limit and chiral symmetry ~\cite{Wise:1992hn,Neubert:1993mb,Burdman:1992gh,Yan:1992gz, Casalbuoni:1992dx,Casalbuoni:1992fd}. In the heavy quark limit, the $S$ wave heavy light mesons are degenerated and could be described in a $H$ matrix, which is,
\begin{eqnarray}
H_{a}^{(Q)}&=&\frac{1+v\!\!\!\slash}{2}[P_{a}^{*(Q)\mu}\gamma_{\mu}-P^{(Q)}_{a}\gamma_{5}].
\end{eqnarray}
As for the $P$-wave heavy-light mesons,  the spin singlet and triplets could be separated into two doublets i.e., $S$ doublet and $T$ doublet, based on light quark degrees of freedom, $\vec s_\ell=\vec s_q +\vec \ell$, which is a conserved quantity in the heavy quark limit. The particular expressions of $S$ and $T$ matrices are defined as,

\begin{eqnarray}
S^{(Q)}_{a}&=&\frac{1+v\!\!\!\slash}{2} \left[P^{'(Q)\mu}_{1a}\gamma_{\mu}\gamma_{5}
        -P^{*(Q)}_{0a} \right]\nonumber\\
T^{(Q)\mu}_{a}&=&\frac{1+v\!\!\!\slash}{2} \left[P^{*(Q)\mu\nu}_{2a}\gamma_{\nu}-
          \sqrt{\frac{3}{2}}P^{(Q)}_{1a\nu}\gamma_{5} \left (g^{\mu\nu}
           -\frac{1}{3}\gamma^{\nu}(\gamma^{\mu}-v^{\mu}) \right) \right ]. \nonumber\\
\end{eqnarray}
As for pion and kaon, since they are considered as chiral particles, the effective interactions between the heavy-light mesons and chiral particles are ,
 \begin{eqnarray}
 \mathcal{L}_{\rm int}&=&ih \left\langle S {A}\!\!\!\slash\gamma_{5} \overline{H}\right\rangle
             +i\frac{h^\prime}{\Lambda_{\chi}} \left\langle T^{\mu}\left (D_{\mu}\ A\!\!\!\slash +D\!\!\!\!\slash \  A_{\mu}\right)\gamma_{5}\overline{H} \right \rangle \nonumber\\ \label{Eq:LagA}
  \end{eqnarray}
where the covariant derivative $D_\mu= \partial_\mu + V_\mu$.  The vector current $V_\mu$ and axial current $A_\mu$ are defined as  
\begin{eqnarray}
V_\mu &=& \frac{1}{2} \left[\xi^\dagger \partial_\mu \xi +\xi \partial_\mu \xi^\dagger \right]\nonumber\\
A_\mu &=& \frac{1}{2} \left[\xi^\dagger \partial_\mu \xi -\xi \partial_\mu \xi^\dagger \right]\nonumber
\end{eqnarray}
with $\xi^2=\mathrm{exp}(2i  {M}/f_{\pi})$  and ${M}$ to be the  chiral particle matrix defined as,
 \begin{eqnarray}
 {M}&=&\left(
               \begin{array}{ccc}
                 \sqrt{\frac{1}{2}}\pi^{0}+ \sqrt{\frac{1}{6}}\eta & \pi^{+} & K^{+} \\
                 \pi^{-} &  -\sqrt{\frac{1}{2}}\pi^{0}+ \sqrt{\frac{1}{6}}\eta & K^{0} \\
                 K^{-} & \overline{K}^{0} & -\sqrt{\frac{2}{3}}\eta \\
               \end{array}
             \right).
 \end{eqnarray}

By expanding the effective interactions in Eq.  (\ref{Eq:LagA}), one can obtain the effective Lagrangians related to $P-$ and $S-$wave charmed mesons and a chiral particle, which are,
\begin{eqnarray}
\mathcal{L}_{P_{0}^{*}PM}&=&ig_{P_{0}^{*}PM}(P^{*}_{0a}
           {\stackrel{\leftrightarrow}{\partial}}_{\mu}P_{b}^{\dag})\partial^{\mu}M+H.c
       \nonumber\\
\mathcal{L}_{P_{1}^{'}P^{*}M}&=&ig_{P_{1}^{'}P^{*}M}(P^{'\mu}
            {\stackrel{\leftrightarrow}{\partial}}_{\nu} P_{\mu}^{*\dagger})\partial^{\nu}M+H.c\nonumber\\
\mathcal{L}_{P_{1}P^{*}M}&=&g_{P_{1}P^{*}M}[3P^{\mu}_{1}(\partial_{\mu}
            \partial_{\nu}M)P^{*\nu\dagger}-P^{\mu}_{1}
            (\partial^{\nu}\partial_{\nu}M)P^{*\dag}_{\mu}]+H.c\nonumber\\
\mathcal{L}_{P^{*}_{2}PM}&=&  g_{P_2^\ast P M}P^{*\mu\nu}_{2}( \partial_{\mu}\partial_{\nu} M)P^{\dagger}+H.c     \nonumber\\
\mathcal{L}_{P^{*}_{2}P^{*}M}&=&ig_{P^{*}_{2}P^{*}M}
            \varepsilon^{\alpha\lambda\eta\sigma}
            (P^{*\mu}_{2\eta}{\stackrel{\leftrightarrow}{\partial}}_\sigma
            P^{*\dag}_{\alpha})(\partial_\mu\partial_{\lambda} M)+H.c.
\end{eqnarray}
In the heavy quark limit and chiral symmetry, all the coupling constants in the above effective Lagrangians can be related to the gauge couplings $h$ or $h^\prime$ by
\begin{eqnarray}
g_{P_{0}^{*}PM}&=&  g_{P_{1}^{'}P^{*}M} =-\frac{h}{f_{\pi}}\nonumber\\
 g_{P_{1}P^{*}M}&=&-2\sqrt{\frac{2}{3}}\frac{h'}{\Lambda_{\chi}f_{\pi}}
            \sqrt{m_{P_{1}} m_{P^{*}}}    \nonumber\\
 g_{P^{*}_{2}P^{*}M}&= &  -\frac{2h^{\prime}}{\Lambda_{\chi}f_{\pi}}\nonumber\\
 g_{P^{*}_{2}PM}&=& \frac{4h^{\prime}}{\Lambda_{\chi}f_{\pi}}\sqrt{m_{P_{2}^{*}}m_{P}}.
 \end{eqnarray}
where $f_\pi=132$ MeV is the decay constant of  pion  and $\Lambda_\chi=1$ GeV is the chiral symmetry-breaking scale. The gauge couplings $h$ and $h^\prime $ are estimated to be   $h=0.56\pm0.04$ and $h^\prime =0.43\pm0.01$, respectively  ~\cite{Ding:2008gr,Colangelo:2012xi}.

The involved effective Lagrangians related to baryons read as ~\cite{Dong:2014ksa},
  \begin{eqnarray}
     \mathcal{L}_{\Lambda_{c}pD}&=&ig_{\Lambda_{c}pD}\bar{\Lambda}_{c}
            \gamma_{5}pD^{0}+H.c\\
     \mathcal{L}_{\Lambda_{c}pD^{\ast0}}&=&g_{\Lambda_{c}pD^{\ast0}}
            \bar{\Lambda}_{c} \gamma^{\mu}pD_{\mu}^{*0}+H.c,
  \end{eqnarray}
where the related coupling constants  are $g_{\Lambda_{c}pD}=-13.98$ and $g_{\Lambda_{c}pD^{\ast0}}=-5.20$ ~\cite{He:2016pfa,Dong:2014ksa}.

With the above effective Lagrangians, we get the amplitudes corresponding to diagrams in Fig.~\ref{Fig:Feyn}, which are,
  \begin{eqnarray}
\mathcal{M}_{a}&=&[g_{\Lambda_{c}pD^{0}}\bar{u}(p_{4})\gamma_{5}
               u(p_{2})]\nonumber\\
           &&\times [-ig_{D^{*}_{s0}D^{0}\pi}(iq^{\mu}+ip_{3}^{\mu})(ip_{1\mu})]
           \frac{1}{q^{2}-m_{D^{0}}^{2}}\nonumber\\
           &&\times F^{2}(q^{2},m_{D^{0}})\nonumber\\
\mathcal{M}
         _{b}&=&[g_{\Lambda_{c}pD^{*0}}\bar{u}(p_{4})
         \gamma_{\beta}
         u(p_{2})]\nonumber\\ &&\times[ g_{D_{1}D^{*0}\pi}(iq_{\nu} ip_{1}^{\nu}
           +ip_{3\nu}ip_{_{1}}^{\nu})\epsilon_{\mu}(p_{3})]\nonumber\\ &&\times
           \frac{-g^{\beta\mu}+q^{\beta} q^{\mu}/m_{D^{\ast0}}^{2}}{q^{2}-m^{2}_{D^{\ast0}}}
           F^{2}(q^{2},m_{D^{\ast0}})\nonumber\\
\mathcal{M}_{c}&=& [g_{\Lambda_{c}pD^{\ast0}}\bar{u}(p_{4})\gamma_{\beta}
         u(p_{2})]\nonumber\\
           &&\times
           [g_{D^{\prime}_{1}D^{\ast0}\pi}(3ip_{1\mu}ip_{1\rho}
           -(ip^{\nu}_{1})(ip_{1\nu})g_{\mu\rho})\epsilon^{\mu}(p_{3})]\nonumber\\ &&\times
           \frac{-g^{\beta\rho}+q^{\beta} q^{\rho}/m_{D^{\ast0}}^{2}}{q^{2}-m^{2}_{D^{\ast0}}}F^{2}
           (q^{2},m_{D^{\ast0}})\nonumber\\
\mathcal{M}_{d}&=&[ig_{\Lambda_{c}pD^{0}}\bar{u}(p_{4})\gamma_{5}u(p_{2})] [g_{D^{\ast}_{2}D^{0}\pi}\epsilon^{\alpha\beta}_{D^{\ast}_{2}}(ip_{1\alpha} ip_{1\beta})]\nonumber\\
           &&\times \frac{1}{q^2-m_{D_0}^2} F^{2}(q^{2},m_{D^{0}})\nonumber\\
\mathcal{M}_{e}&=&[g_{\Lambda_{c}pD^{\ast0}}\bar{u}(p_{4})
           \gamma _{\beta} u(p_{2}) ]\nonumber\\
           && \times [ -ig_{D_{2}^{\ast}D^{\ast0}\pi}\varepsilon_{\alpha\lambda\eta\sigma}
\epsilon_{D^{\ast}_{2}}^{\zeta\eta}(iq^\sigma+ip_{3}^\sigma) (ip^{\zeta}_{1 } ip^{\lambda}_{1})]\nonumber\\ &&\times \frac{-g^{\beta\alpha}+q^{\beta} q^{\alpha}/m_{D^{\ast0}}^{2}}{q^{2}-m^{2}_{D^{\ast0}}}F^{2} (q^{2},m_{D^{\ast0}}),
\end{eqnarray}
where $F (q, m_{D^{(\ast)}})$ is the form factor introduced to depict the off-shell effects of the exchanged mesons as well as the structure effects of the involved mesons.  The specific expression of the form factors will be discussed in the following section.

With the above amplitudes, we can estimate the differential cross sections of the pion induced $P$-wave charmed meson production process, which is
\begin{eqnarray}
      \frac{d\sigma}{d\cos\theta}=\frac{1}{32\pi s}\frac{| \vec{p}_{f}|}{|\vec{p}_{i}|} \left(\frac{1}{2}\overline{|\mathcal{M}|^{2}}\right)
\end{eqnarray}
where $s=(p_{1}+p_{2})^{2}$ is the center-of-mass energy, and $\theta$ denotes the angle of the outgoing charmed meson relative to the pion beam direction in the center-of-mass frame. $\vec{p}_{i}$ and $\vec{p}_{f}$ are the  three-momenta of the initial pion beam and final charmed meson, respectively. The factor $1/2$ in the bracket comes from the spin average of the initial state and the overline indicates the sum over the final state spins.

\section{Numerical Results and discussions}
\label{Sec:Num}

In the present work, we adopt the form factor in a monopole form, which is~\cite{Chen:2013cpa,Chen:2014ccr}
\begin{eqnarray}
    F\left(q^{2},m^{2}_{D^{(\ast)}}\right)=
    \frac{m^{2}_{D^{(\ast)}}-\Lambda_{D^{(\ast)}}^{2}}{q^{2}
    -\Lambda^{2}_{D^{(\ast)}}}
    \label{Eq:FFs1}
\end{eqnarray}
where the parameter $\Lambda$ can be further reparameterized as $\Lambda_{D^{(\ast)}}=m_{D^{(\ast)}}+\alpha\Lambda_{\rm QCD} $ with
   $\Lambda_{\rm QCD}=0.22 \ {\rm GeV}$ and $m_{D^{(\ast)}}$ is the mass of the exchanged meson. The model parameter $\alpha$ should be of order of unity~\cite{Tornqvist:1993vu,Tornqvist:1993ng,Locher:1993cc,Li:1996yn} but its concrete value cannot be estimated by the first principle. In practice, the value of $\alpha$ is usually determined by comparing theoretical estimates with the corresponding experimental measurements. Unfortunately, the experimental data for the involved processes are not available now. So in the present work, we adopt the same $\alpha$ values for all the $P-$ wave charm/charmed-strange meson production processes due to the similarity of the involved mesons by varying this model parameter around unity. To further reduce the uncertainties caused by the present model, we focus on the relative size of the cross sections for the different final states, which should be weakly dependent on the form factors. Further discussions about form factors can be found in Appendix~\ref{App:FFs}.

\subsection{Cross sections for $P-$wave charmed and charmed strange meson production}

\begin{figure}[htb]
  \includegraphics[width=8cm]{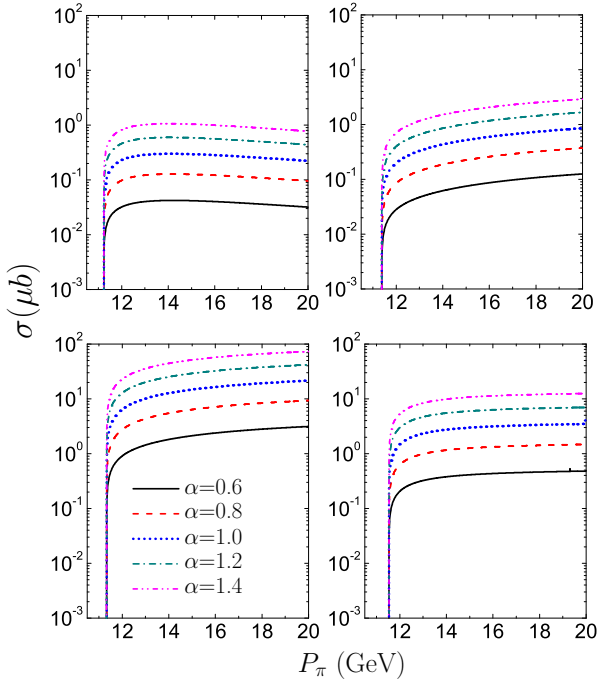}
  \caption{The cross sections for $P-$wave charmed meson production processes depending on the momentum of  pion beam. Diagrams $(a)$, $(b)$, $(c)$, and $(d)$ are cross sections for $D_{0}^\ast(2400)$, $D_1^\prime (2430)$, $D_1(2420)$, and $D_2^\ast (2460)$ productions, respectively. \label{Fig:CS-a} }
\end{figure}

The cross sections for $P-$wave charmed meson production by a pion induced reaction on a proton target are presented in Fig.~\ref{Fig:CS-a}. Taking the process $\pi^- p \to D_0^{\ast -} (2400) \Lambda_c^{+}$ [Fig.~\ref{Fig:CS-a} (a)] as an example, the cross section sharply increases near the threshold of $D_0^{\ast -}(2400) \Lambda_c^{+}$. However, with the beam momentum up to 13 GeV, the cross section becomes very weakly dependent on the beam momentum. In particular, the cross sections are about $2 \times 10^{-2}\ \mu b$ for $\alpha=0.6$ and about $1\ \mu b$ for $\alpha=1.4$, which indicates the cross sections strongly depend on the model parameter $\alpha$. As for $\pi p \to D_2^\ast \Lambda_c$ [Fig.~\ref{Fig:CS-a}-(d)], the $P_\pi$ dependences of the cross sections are very similar to those of $\pi p \to D_0^\ast (2400) \Lambda_c$ but different in magnitude, which is about one order larger than those of $D_0^\ast (2400)$ production process.  As for $\pi p \to D_1^\prime (2430) \Lambda_c$ [Fig.~\ref{Fig:CS-a}-(b)] and $\pi p \to D_1(2420) \Lambda_c$ [Fig.~\ref{Fig:CS-a}-(c)], their cross sections still increase slowly with $P_\pi$ increasing after the shape rise near the threshold.

\begin{figure}[t]
\includegraphics[width=8cm]{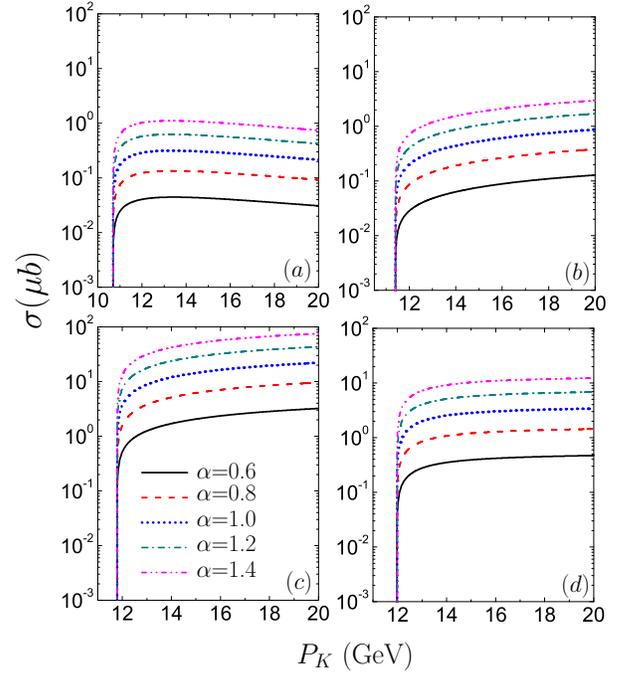}
\caption{The cross sections for $P-$wave charmed strange meson production processes depending on the momentum of  kaon beam. Diagrams $(a)$, $(b)$, $(c)$, and $(d)$ are cross sections for $D_{s0}^\ast(2317)$, $D_{s1}^{\prime} (2460)$, $D_{s1}(2536)$, and $D_{s2}^\ast (2573)$ productions, respectively. \label{Fig:CS-b} }
\end{figure}

The cross sections for $P-$wave charmed-strange meson production by a kaon induced reaction on a proton target are presented in Fig.~\ref{Fig:CS-b}. The behaviors of the cross sections are very similar to those of charmed meson production processes. After the sharp rise near the threshold, the cross sections of $D_{s0}^\ast(2317)$ and $D_{s2}(2573)$ production processes become very weakly dependent on the kaon beam momentum, while those of $D_{s1}^\prime(2460)$ and $D_{s1}(2536)$ production processes still increase slowly.  The estimations indicate the cross sections for these processes are also strongly dependent on the model parameter $\alpha$.

\begin{figure}[hbt]
  \includegraphics[width=8cm]{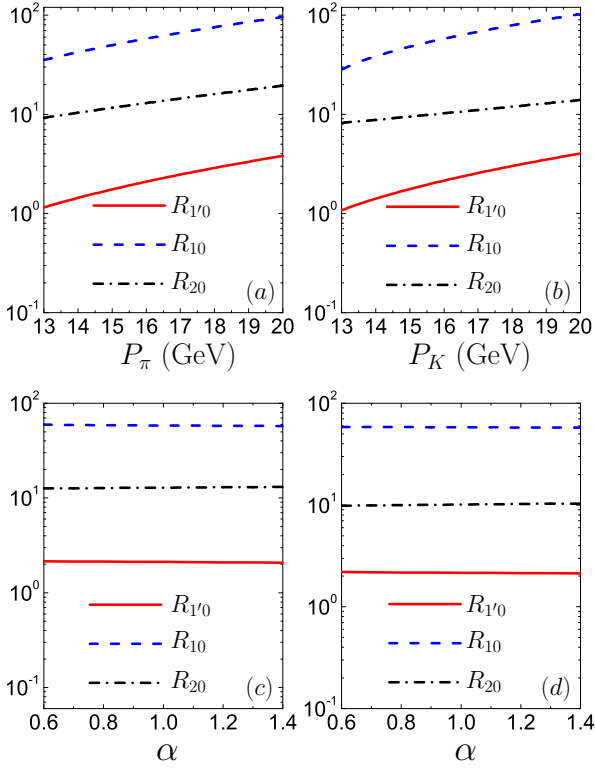}
  \caption{The cross section ratios for $P-$wave charmed (left panel) and charmed-strange (right panel) meson productions depending on beam momentum (upper panel)and the model parameter $\alpha$ (lower panel). \label{Fig:Ratio} }
\end{figure}

To further reduce the uncertainties resulted from the model parameter $\alpha$,  taking a $\pi p$ reaction as an example, we define the following cross section ratios,
\begin{eqnarray}
R_{1^\prime 0} &=& \frac{\sigma(\pi p \to D_1^\prime(2430) \Lambda_c)}{\sigma(\pi p \to D_0^\ast(2400) \Lambda_c)} \nonumber \\
R_{1 0} &=& \frac{\sigma(\pi p \to D_1(2420) \Lambda_c)}{\sigma(\pi p \to D_0^\ast(2400) \Lambda_c)} \nonumber \\
R_{2 0} &=& \frac{\sigma(\pi p \to D_2^\ast (2460) \Lambda_c)}{\sigma(\pi p \to D_0^\ast(2400) \Lambda_c)}.
\end{eqnarray}
In the same way, we can define the cross section ratios for a $Kp$ reaction processes.

In Fig.~\ref{Fig:Ratio}, we present the ratios defined above depending on the beam momentum and model parameter. In particular, for the $\pi p$ reaction processes with $\alpha=1.2$ [as shown in Fig.~\ref{Fig:Ratio}-(a], the ratio $R_{1^\prime 0}$ is about $1.14 \sim 3.94 $  when $P_\pi= (13 \sim 20) \ {\rm GeV}$. As for $R_{10}$, the ratio is estimated to be $35.05\sim 98.35$, which indicates the cross sections for $\pi p \to D_1(2420) \Lambda_c$ is about 30 times of the one for  $\pi p \to D_1^\prime (2430) \Lambda_c$. In the heavy quark limit, $D_1^\prime(2430)$ and $D_1 (2420)$ couples to $D^\ast \pi$ via $S-$wave and $D-$wave, respectively. Accordingly, we normally expect that the width of $D_1^\prime (2430)$ is much larger than the one of $D_1(2420)$.  The low partial wave coupling will be enhanced in the low energy scales, i.e., the involved momentum is very small, such as the decay processes of $D_1^{\prime} (2430)$ and $D_1 ({2420})$. However, in the production process, the momentum of pion is very large. In this case, the $D$ wave coupling will be enhanced since the vertex $D_1(2420) D^\ast \pi \sim p_\pi^2$, while $D_1(2430) D^\ast \pi \sim p_\pi \cdot p_{D^\ast}$. As for $R_{20}$, it weakly depends on the beam momentum and varies from 9.13 to 19.66 in the considered beam momentum range.  As shown in Fig.~\ref{Fig:Ratio}-(b), the ratios of $Kp $ reaction processes are very similar to those of $\pi p $ reaction processes. To further test the model dependent of the ratios, we present the $\alpha$ dependences of these ratios in Fig.~\ref{Fig:Ratio}-(c) and (d) with $P_{\pi/K}=16\ {\rm GeV}$. Our estimations indicate that these cross sections ratios are almost independent on the model parameter $\alpha$ and the effects of the form factors can be almost completely canceled in the ratios, which are resulted from the similarities of kinematics for considered processes.

\begin{figure}[t]
\includegraphics[width=8cm]{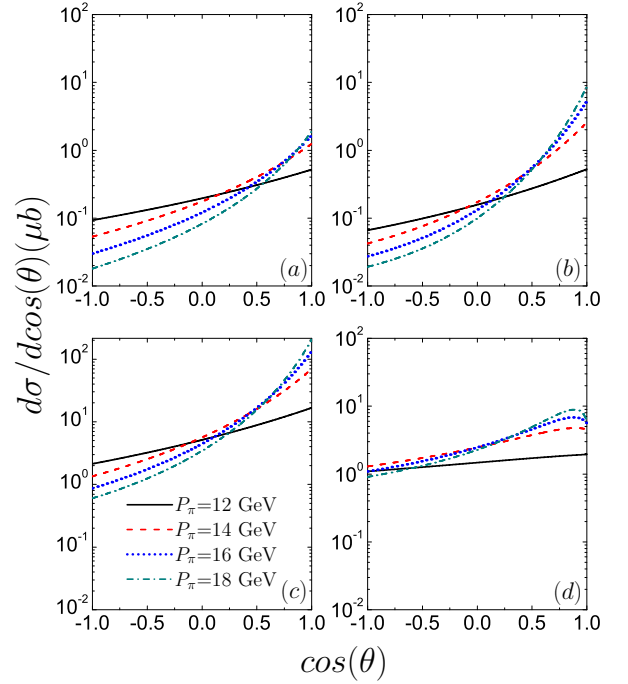}
 \caption{The same as Fig.~\ref{Fig:CS-a} but for differential cross sections depending on $\cos \theta$ \label{Fig:dCS-a}}
\end{figure}

\begin{figure}[htb]
\includegraphics[width=8cm]{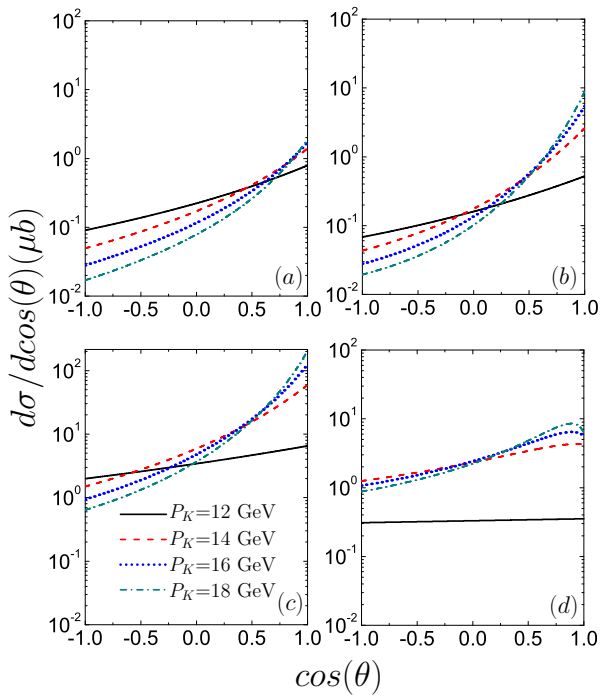}
 \caption{The same as Fig.~\ref{Fig:CS-b} but for differential cross sections depending on $\cos \theta$ \label{Fig:dCS-b} }
\end{figure}

\subsection{Differential cross sections}
Besides the cross sections, we present the differential cross sections of the considered processes depending on $\cos \theta$, where $\theta$ is the angle between the outgoing charmed or charmed-strange meson and the beam direction. Since our estimates indicate that the shape of differential cross sections weakly depend on the model parameter, we fix the model parameter $\alpha= 1.2 $ in the represent work to see the $\cos \theta$ distributions at different beam momenta, which are 12, 14, 16 , and 18 GeV for Fig.~\ref{Fig:dCS-a}-(a)$\sim$(d), respectively. In Fig.~\ref{Fig:dCS-a}, we present the $\cos \theta$ distributions of $P-$wave charmed meson productions in the $\pi p$ reaction process. Here, one should notice that the differential cross sections reach the maximum at the forward angle limit. As $P_\pi$ increases, more charmed mesons are concentrated in the forward angle area. As for the $D_2^\ast (2460)$ production process, the threshold is  $11.53$ GeV, which is close to 12 GeV. Thus, the produced $D_2^\ast(2460)$ has small momentum and the distributions of the differential cross sections weakly depend on $\cos \theta$ as seen in Fig.~\ref{Fig:dCS-a}-(d).

The differential cross sections for $P$-wave charmed-strange meson productions are presented in Fig.~\ref{Fig:dCS-b}.  The differential cross sections increases as $\cos (\theta)$ increases, which is the same as the $P-$wave charmed meson production processes.  As for the $D_{s2}^\ast (2573)$ production process, the threshold is $11.98$ GeV, which is extremely close to 12 GeV. Thus, the differential cross sections are almost independent on $\cos \theta$ and much smaller than those with higher beam momenta as seen in Fig.~\ref{Fig:dCS-b}-(d).

\section{Summary}
\label{Sec:Sum}
In the present work, we have evaluated the production of $P-$wave charmed/charmed strange mesons in the pion/kaon induced reactions on a proton target by an effective Lagrangian approach. The cross sections as well as the differential cross sections depending on the scattering angle have been evaluated. Our estimation indicates that the cross sections for the considered processes shapely increase near the threshold and weakly depend on the incoming beam momentum. For both charmed and charmed-strange mesons, the cross sections for the $S$ doublet are of the same order, while cross sections for the $T$ doublet are about one order larger than those for the $S$ doublet.

Our predictions for the cross sections are not convincing as we have expected since the magnitude of cross sections strongly depend on the model parameter. However, we have found such model dependence completely disappears in the cross section ratios, which is the best out of the worst. The cross section ratios for the charmed meson production processes provide us an opportunity to check the validity of heavy quark limit in the charmed region, and the ratios for the charmed-strange meson production processes can further test our understanding of  $P-$wave charmed-strange mesons, especially for $D_{s0}^\ast(2317)$ and $D_{s1}(2460)$.

\section* {Acknowledgements}
This project is supported by the National Natural Science Foundation of China (Grants No. 11775050, and No. 11675228), the Major State Basic Research Development Program in China under grant 2014CB845405, and the Fundamental Research Funds for the Central Universities.

\appendix

\begin{figure}[htb]
\includegraphics[width=8cm]{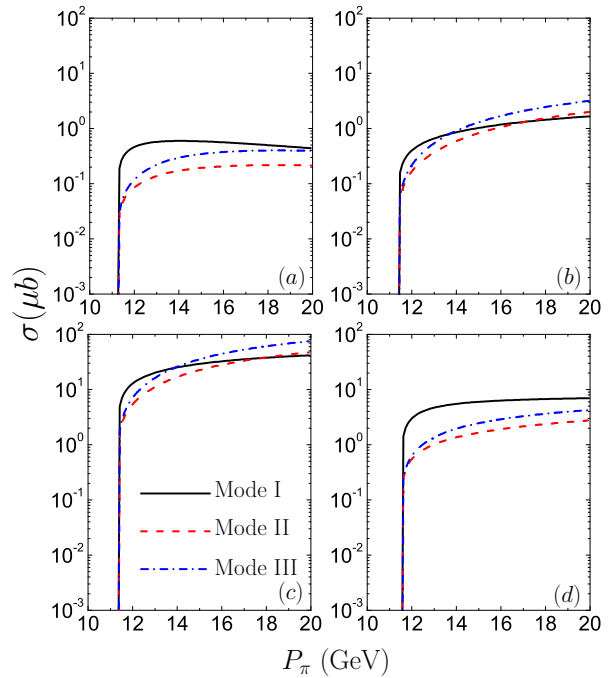}
 \caption{The cross sections for $P-$wave charmed meson production processes with different form factors depending on the momentum of pion beam. Here the parameter $\alpha$ is taken an 1.2, -1.6 and -0.4 in Mode I, Mode II and Mode III, respectively. \label{Fig:App-CS-a} }
\end{figure}

\begin{figure}[htb]
\includegraphics[width=8cm]{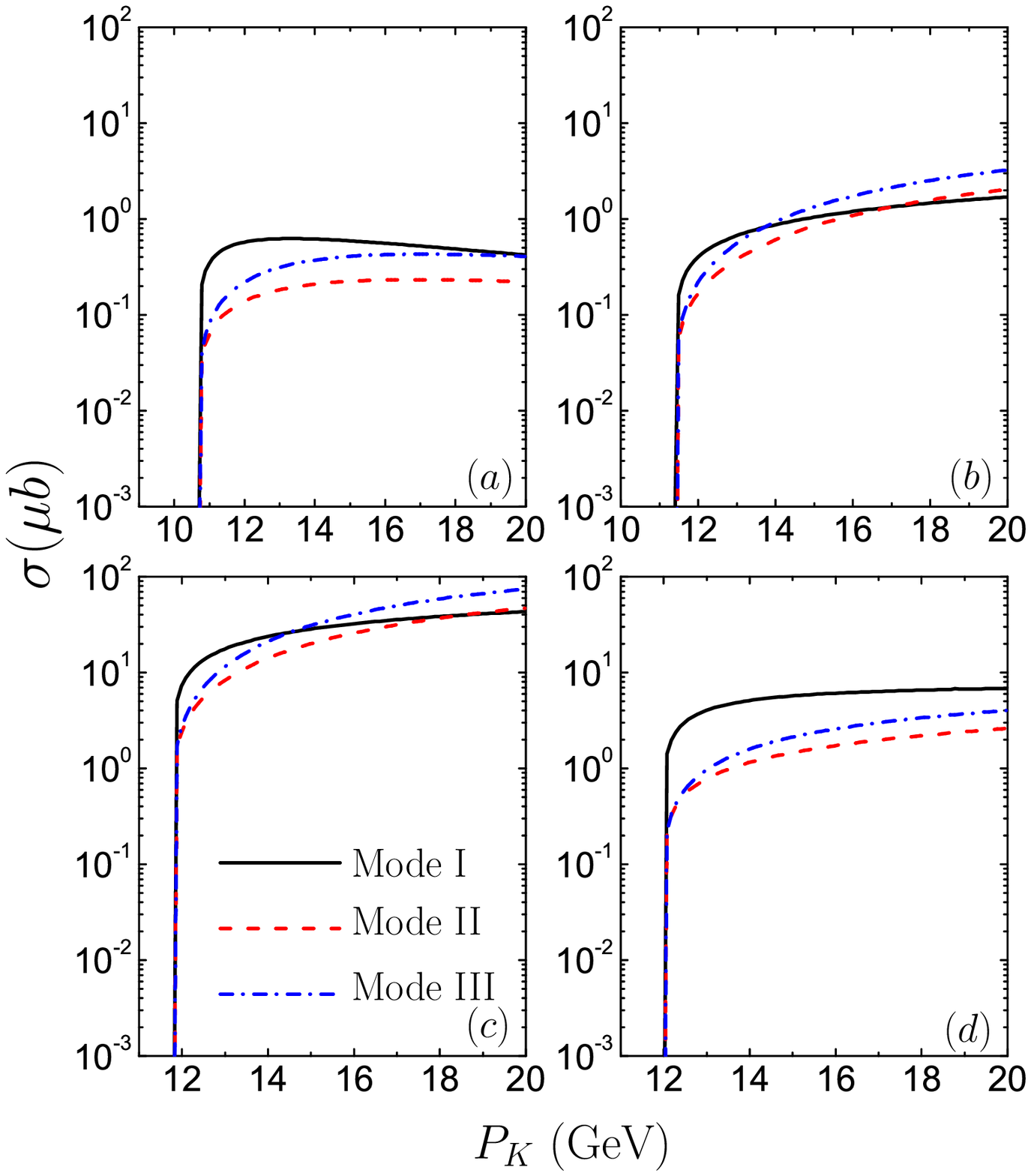}
 \caption{The same as Fig.~\ref{Fig:App-CS-a} but for $P-$wave charmed-strange mesons productions. \label{Fig:App-CS-b} }
\end{figure}

\begin{figure}[htb]
\includegraphics[width=8cm]{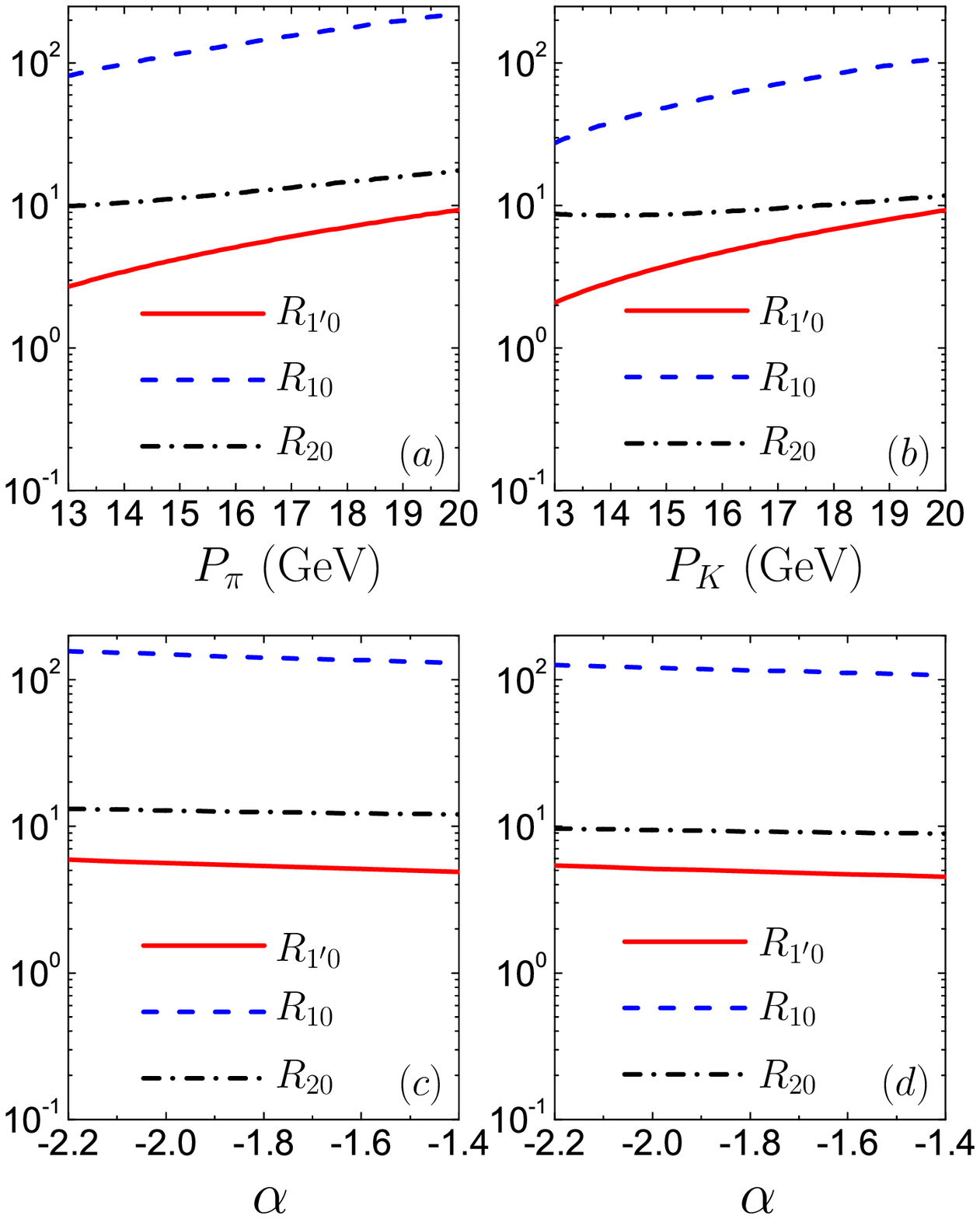}
 \caption{The same as Fig.~\ref{Fig:Ratio} but for Mode II. \label{Fig:App-Ratio2} }
\end{figure}

\begin{figure}[htb]
\includegraphics[width=8cm]{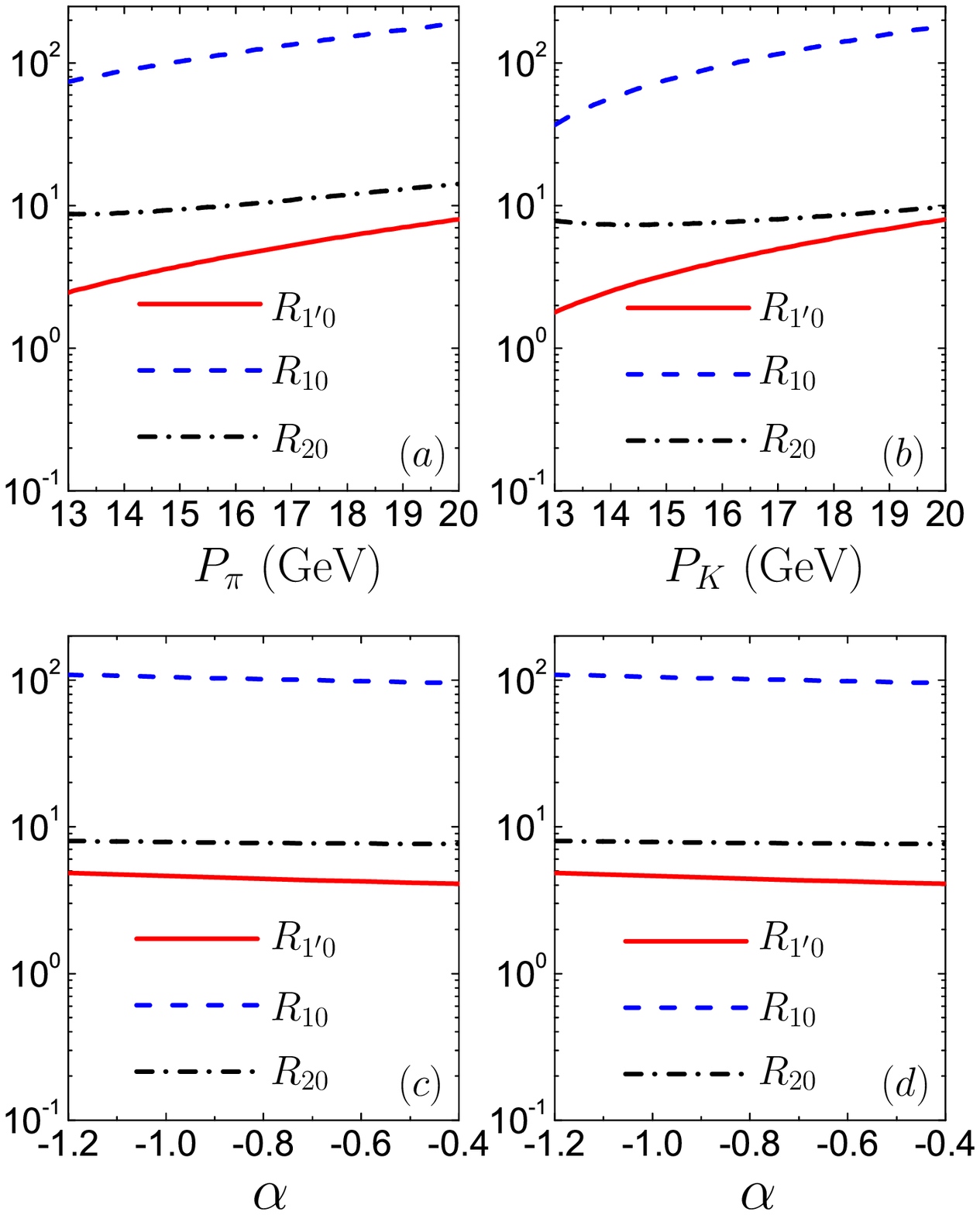}
 \caption{The same as Fig.~\ref{Fig:Ratio} but for Mode III. \label{Fig:App-Ratio3} }
\end{figure}

\section{Comparisons of different form factors}
\label{App:FFs}
As we clarified in Section~\ref{Sec:Num}, the form factors and parameters can usually be determined by comparing the theoretical estimations and the related experimental data. However, as for the processes considered in the present work, there is no experimental data available till now. Thus, we use a form factor in monopole form with the parameter $\alpha$ around the unity according to previous experiences~\cite{Tornqvist:1993vu, Tornqvist:1993ng, Locher:1993cc, Li:1996yn}. Actually, the form factor could be different form, such as~\cite{He:2019rva},
\begin{eqnarray}
	F\left(q^2,m^2_{D^(\ast)}\right) &=& \frac{\Lambda^4}{(m^2_{D^{(\ast)}}-q^2)^2+\Lambda^4},\\
	F\left(q^2,m^2_{D^(\ast)}\right) &=& \mathrm{exp}\left(-(m^2_{D^{(\ast)}}-q^2)/\Lambda^2\right).	
\end{eqnarray}
Here, we take the form factors in above two forms as Mode II and Mode III,respectively, while the one in Eq.~(\ref{Eq:FFs1}) as Mode I.For comparison, we also parametrized the parameter $\Lambda$ in the above equations as $\Lambda=m_{D^{(\ast)}} +\alpha \Lambda_{QCD}$. However, it should be noticed that the $\alpha$ in different form factors are different if they are used to depict the same processes. For example, in the first type of form factor $\alpha$ should be larger than $0$ to avoid an unphysical suppression at $\Lambda=m_{D^{(\ast)}}$, while in other two types form factors, $\alpha$ could be zero. Thus, in this section, we first try to find a proper $\alpha$ for different form factors by reproducing a similar cross sections for the processes considered in the present work and then check the cross section ratios and the differential cross sections. We find that when taking $\alpha=-1.6$ for Mode II and $\alpha=-0.4$ for Mode III, we can produce a similar cross sections as those for Mode I with $\alpha=1.2$. The cross sections depending on the beam momentum with different form factors are present in Figs.~\ref{Fig:App-CS-a} and \ref{Fig:App-CS-b}.

With the parameters determined by the cross sections for Mode II and Mode III, we can further check the cross sections ratios for the processes involved in the present work, which are presented in Figs.~\ref{Fig:App-Ratio2} and \ref{Fig:App-Ratio3}. From these figures one can find the beam momentum and $alpha$ dependences of the cross sections ratios are very similar for different form factors. Taking Mode II as an example, for the charmed mesons productions, the ratio $R_{1^\prime 0}$, $R_{10}$ and $R_{20}$ are $2.69\sim9.33$, $81.31\sim222.93$ and $9.92\sim17.58$, respectively, and for the charmed strange mesons productions, these ratios are estimated to be, $2.08\sim9.28$, $27.45\sim110.35$ and $8.75\sim11.75$, respectively. The $\alpha$ dependences of the cross sections ratios in Mode II and Mode III as shown in Figs.~\ref{Fig:App-Ratio2}-(c,d) and \ref{Fig:App-Ratio3}-(c,d), respectively, indicate that the cross section ratios are almost independent on the model parameter, which are the same as the one in Mode I.

\begin{figure}[htb]
\includegraphics[width=8cm]{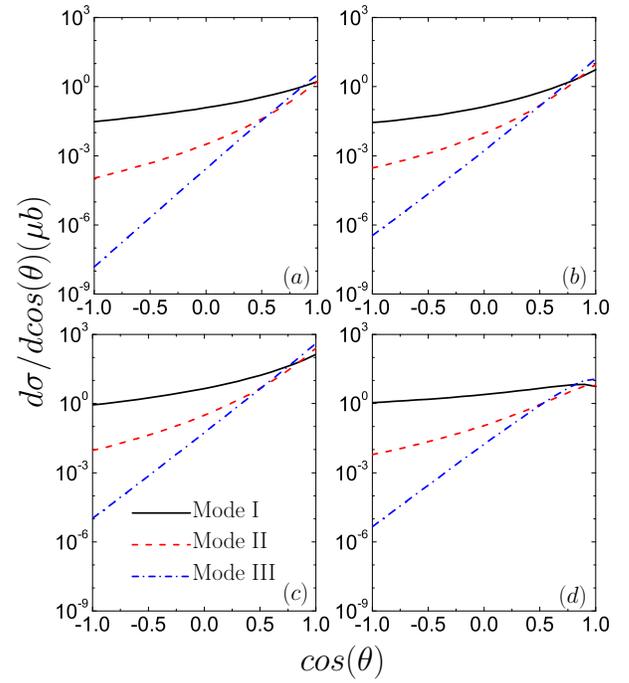}
 \caption{The same as Fig.~\ref{Fig:App-CS-a} but for differential cross sections depending on $\cos \theta$ \label{Fig:App-dCS-a} }
\end{figure}

\begin{figure}[htb]
\includegraphics[width=8cm]{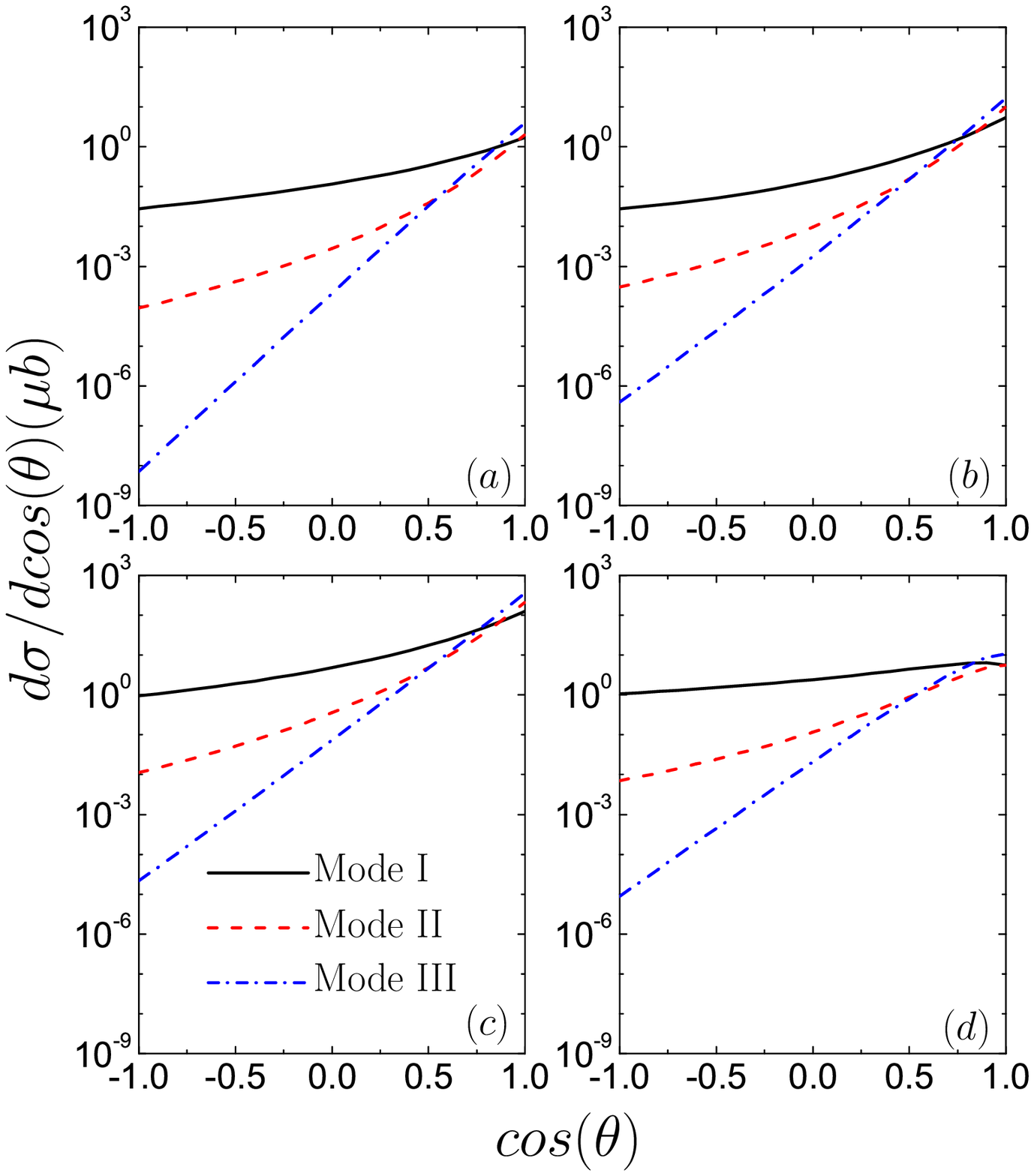}
 \caption{The same as Fig.~\ref{Fig:App-dCS-a} but for $P-$wave charmed-strange mesons productions. \label{Fig:App-dCS-b} }
\end{figure}

The differential cross sections for the $P-$ wave charmed and charmed strange meson productions in Mode II and Mode III are presented in Fig.~\ref{Fig:App-dCS-a} and \ref{Fig:App-dCS-b}, respectively. From the figure, one can find the differential cross sections reach their maximum at the forward angle limit and in the backward angle limit, the differential cross sections could be ignored,  which are the same as the cases in Model I.

\begin{thebibliography}{00}


\bibitem{Godfrey:1985xj}
  S.~Godfrey and N.~Isgur,
  Mesons in a Relativized Quark Model with Chromodynamics,
  Phys.\ Rev.\ D {\bf 32}, 189 (1985).

\bibitem{Albrecht:1985as}
  H.~Albrecht {\it et al.} [ARGUS Collaboration],
  Observation of a New Charmed Meson.,
  Phys.\ Rev.\ Lett.\  {\bf 56} (1986) 549.

\bibitem{Albrecht:1988dj}
  H.~Albrecht {\it et al.} [ARGUS Collaboration],
  Observation of the $D^*_{0}$ (2459) in $e^+ e^-$ Annihilation,
  Phys.\ Lett.\ B {\bf 221}, 422 (1989).


\bibitem{Link:2003bd}
  J.~M.~Link {\it et al.} [FOCUS Collaboration],
  Measurement of masses and widths of excited charm mesons $D_2^*$ and evidence for broad states,
  Phys.\ Lett.\ B {\bf 586}, 11 (2004)


\bibitem{Abe:2003zm}
  K.~Abe {\it et al.} [Belle Collaboration],
  Study of $B^- \to D^{\ast \ast 0} \pi^- (D^{\ast\ast 0 } \to  D^{(\ast)+}\pi^-)$ decays,
  Phys.\ Rev.\ D {\bf 69} (2004) 112002

\bibitem{Asratian:1987rb}
  A.~E.~Asratian {\it et al.},
  Studying (Anti-charm Strange) Spectroscopy in Anti-neutrino $N$ Collisions,
  Z.\ Phys.\ C {\bf 40}, 483 (1988).


\bibitem{Aubert:2003fg}
  B.~Aubert {\it et al.} [BaBar Collaboration],
  Observation of a narrow meson decaying to $D_s^+ \pi^0$ at a mass of 2.32-GeV/c$^2$,
  Phys.\ Rev.\ Lett.\  {\bf 90}, 242001 (2003)


\bibitem{Besson:2003cp}
  D.~Besson {\it et al.} [CLEO Collaboration],
  Observation of a narrow resonance of mass 2.46 GeV/c$^2$ decaying to $D^{*+}_s \pi^0$ and confirmation of the $D^*_{sJ}(2317)$ state,
  Phys.\ Rev.\ D {\bf 68} (2003) 032002
   Erratum: [Phys.\ Rev.\ D {\bf 75} (2007) 119908]

\bibitem{Xie:2010zza}
  Z.~X.~Xie, G.~Q.~Feng and X.~H.~Guo,
  Analyzing $D_{s0}^*(2317)^+$ in the $DK$ molecule picture in the Beth-Salpeter approach,
  Phys.\ Rev.\ D {\bf 81},  036014 (2010).


\bibitem{Zhang:2006ix}
  Y.~J.~Zhang, H.~C.~Chiang, P.~N.~Shen and B.~S.~Zou,
  Possible $S$-wave bound-states of two pseudoscalar mesons,
  Phys.\ Rev.\ D {\bf 74}, 014013 (2006)


\bibitem{Bicudo:2004dx}
  P.~Bicudo,
  The Family of strange multiquarks as kaonic molecules bound by hard core attraction,
  Nucl.\ Phys.\ A {\bf 748},  537 (2005).


\bibitem{Faessler:2007gv}
  A.~Faessler, T.~Gutsche, V.~E.~Lyubovitskij and Y.~L.~Ma,
  Strong and radiative decays of the $D_{s0}^*(2317)$ meson in the $DK$-molecule picture,
  Phys.\ Rev.\ D {\bf 76}, 014005 (2007).

\bibitem{Faessler:2007us}
  A.~Faessler, T.~Gutsche, V.~E.~Lyubovitskij and Y.~L.~Ma,
  $D^* K$ molecular structure of the $D_{s1}(2460)$ meson,
  Phys.\ Rev.\ D {\bf 76}, 114008 (2007).


\bibitem{Cleven:2014oka}
  M.~Cleven, H.~W.~Grießhammer, F.~K.~Guo, C.~Hanhart and U.~G.~Meißner,
  Strong and radiative decays of the $D^*_{s0}(2317)$ and $D_{s1}(2460)$,
  Eur.\ Phys.\ J.\ A {\bf 50}, no. 9, 149 (2014).

\bibitem{Xiao:2016hoa}
  C.~J.~Xiao, D.~Y.~Chen and Y.~L.~Ma,
  Radiative and pionic transitions from the $D_{s1}(2460)$ to the $D_{s0}^\ast(2317)$,
  Phys.\ Rev.\ D {\bf 93}, no. 9, 094011 (2016)

\bibitem{Datta:2003re}
  A.~Datta and P.~J.~O'donnell,
  Understanding the nature of $D_s(2317)$ and $D_s(2460)$ through nonleptonic B decays,
  Phys.\ Lett.\ B {\bf 572} , 164 (2003).

\bibitem{Lutz:2008zz}
  M.~F.~M.~Lutz and M.~Soyeur,
  Open-charm meson systems in the hadrogenesis conjecture,
  Prog.\ Part.\ Nucl.\ Phys.\  {\bf 61}, 155 (2008).

\bibitem{Hwang:2004cd}
  D.~S.~Hwang and D.~W.~Kim,
  Mass of $D^*_{sJ}(2317)$ and coupled channel effect,
  Phys.\ Lett.\ B {\bf 601} , 137 (2004).

\bibitem{Song:2015nia}
  Q.~T.~Song, D.~Y.~Chen, X.~Liu and T.~Matsuki,
  Charmed-strange mesons revisited: mass spectra and strong decays,
  Phys.\ Rev.\ D {\bf 91}, 054031 (2015)

\bibitem{Liu:2013maa}
  J.~B.~Liu and M.~Z.~Yang,
  Spectrum of the charmed and b-flavored mesons in the relativistic potential model,
  JHEP {\bf 1407}, 106 (2014).

\bibitem{Liu:2006jx}
  X.~Liu, Y.~M.~Yu, S.~M.~Zhao and X.~Q.~Li,
  Study on decays of $D^*_{sJ}(2317)$ and $D_{sJ}(2460)$ in terms of the CQM model,
  Eur.\ Phys.\ J.\ C {\bf 47}, 445 (2006).

\bibitem{Lu:2006ry}
  J.~Lu, X.~L.~Chen, W.~Z.~Deng and S.~L.~Zhu,
  Pionic decays of $D_{sJ}(2317)$, $D_{sJ}(2460)$ and $B_{sJ}(5718)$, $B_{sJ}(5765)$,
  Phys.\ Rev.\ D {\bf 73}, 054012 (2006).



\bibitem{Wang:2006mf}
  Z.~G.~Wang,
  Radiative decays of the $D_{s0}(2317)$, $D_{s1}(2460)$ and the related strong coupling constants,
  Phys.\ Rev.\ D {\bf 75},  034013 (2007).


\bibitem{Fajfer:2015zma}
  S.~Fajfer and A.~P.~Brdnik,
  Chiral loops in the isospin violating decays of $D_{s1}(2460)^+$ and $D^*_{s0}(2317)^+$,
  Phys.\ Rev.\ D {\bf 92} , 074047 (2015).


\bibitem{Dai:2003yg}
  Y.~B.~Dai, C.~S.~Huang, C.~Liu and S.~L.~Zhu,
  Understanding the $D^+_{sJ}(2317)$ and $D^+_{sJ}(2460)$ with sum rules in HQET,
  Phys.\ Rev.\ D {\bf 68} , 114011 (2003).

\bibitem{Colangelo:2003vg}
  P.~Colangelo and F.~De Fazio,
  Understanding $D_{sJ}(2317)$,
  Phys.\ Lett.\ B {\bf 570}, 180 (2003).

\bibitem{Colangelo:2005hv}
  P.~Colangelo, F.~De Fazio and A.~Ozpineci,
  Radiative transitions of $D^*_{sJ}(2317)$ and $D_{sJ}(2460)$,
  Phys.\ Rev.\ D {\bf 72} (2005) 074004.


\bibitem{Nagae:2008zz}
  T.~Nagae,
  The J-PARC project,
  Nucl.\ Phys.\ A {\bf 805}, 486 (2008).

\bibitem{Nerling:2012er} F.~Nerling [COMPASS Collaboration],
 Hadron Spectroscopy with COMPASS: Newest
Results, EPJ Web Conf.\ \textbf{37} (2012) 01016.

\bibitem{Obraztsov:2016lhp}
  V.~Obraztsov [OKA Collaboration],
  High statistics measurement of the $K^+ \to \pi^0 e^+\nu$(Ke3) decay formfactors,
  Nucl.\ Part.\ Phys.\ Proc.\  {\bf 273-275}, 1330 (2016).



\bibitem{Velghe:2016jjw}
  B.~Velghe [NA62-RK and NA48/2 Collaborations],
  $K^{\pm} \to \pi^{\pm} \gamma \gamma$ Studies at NA48/2 and NA62-RK Experiments at CERN,
  Nucl.\ Part.\ Phys.\ Proc.\  {\bf 273-275}, 2720 (2016).


\bibitem{Wise:1992hn}
  M.~B.~Wise,
  Chiral perturbation theory for hadrons containing a heavy quark,
  Phys.\ Rev.\ D {\bf 45}, no. 7, R2188 (1992).



\bibitem{Burdman:1992gh}
  G.~Burdman and J.~F.~Donoghue,
  Union of chiral and heavy quark symmetries,
  Phys.\ Lett.\ B {\bf 280}, 287 (1992).


\bibitem{Casalbuoni:1992dx}
  R.~Casalbuoni, A.~Deandrea, N.~Di Bartolomeo, R.~Gatto, F.~Feruglio and G.~Nardulli,
  Effective Lagrangian for heavy and light mesons: Semileptonic decays,
  Phys.\ Lett.\ B {\bf 299}, 139 (1993)


\bibitem{Yan:1992gz}
  T.~M.~Yan, H.~Y.~Cheng, C.~Y.~Cheung, G.~L.~Lin, Y.~C.~Lin and H.~L.~Yu,
  Heavy quark symmetry and chiral dynamics,''
  Phys.\ Rev.\ D {\bf 46}, 1148 (1992)
  Erratum: [Phys.\ Rev.\ D {\bf 55}, 5851 (1997)].




\bibitem{Neubert:1993mb}
  M.~Neubert,
  Heavy quark symmetry,
  Phys.\ Rept.\  {\bf 245}, 259 (1994)



\bibitem{Casalbuoni:1992fd}
  R.~Casalbuoni, A.~Deandrea, N.~Di Bartolomeo, R.~Gatto, F.~Feruglio and G.~Nardulli,
  Hadronic transitions among quarkonium states in a soft exchange approximation. Chiral breaking and spin symmetry breaking processes,
  Phys.\ Lett.\ B {\bf 309}, 163 (1993)






\bibitem{Ding:2008gr}
  G.~J.~Ding,
  Are $Y(4260)$ and $Z^+_2(4250)$ are $D_1 D$ or $D_0 D^*$ Hadronic Molecules?,
  Phys.\ Rev.\ D {\bf 79}, 014001 (2009)


\bibitem{Colangelo:2012xi}
  P.~Colangelo, F.~De Fazio, F.~Giannuzzi and S.~Nicotri,
  New meson spectroscopy with open charm and beauty,
  Phys.\ Rev.\ D {\bf 86}, 054024 (2012)


\bibitem{Dong:2014ksa}
  Y.~Dong, A.~Faessler, T.~Gutsche and V.~E.~Lyubovitskij,
  Role of the hadron molecule $\Lambda_c$(2940) in the $p\bar{p} \to p D^0\bar{\Lambda}_c$(2286) annihilation reaction,''
  Phys.\ Rev.\ D {\bf 90}, no. 9, 094001 (2014)


\bibitem{He:2016pfa}
  J.~He,
  Understanding spin parity of $P_c(4450)$ and $Y(4274)$ in a hadronic molecular state picture,
  Phys.\ Rev.\ D {\bf 95}, no. 7, 074004 (2017)


 

\bibitem{Chen:2013cpa}
  D.~Y.~Chen, X.~Liu and T.~Matsuki,
  Anomalous radiative transitions between $h_b(nP)$ and $\eta_b(mS)$ and hadronic loop effect,
  Phys.\ Rev.\ D {\bf 87}, no. 9, 094010 (2013)


\bibitem{Chen:2014ccr}
  D.~Y.~Chen, X.~Liu and T.~Matsuki,
  Explaining the anomalous $\Upsilon(5S)\to \chi_{bJ}\omega$ decays through the hadronic loop effect,
  Phys.\ Rev.\ D {\bf 90}, no. 3, 034019 (2014)





\bibitem{Tornqvist:1993vu}
  N.~A.~Tornqvist,
 On deusons or deuteron - like meson meson bound states,
  Nuovo Cim.\ A {\bf 107}, 2471 (1994)


\bibitem{Tornqvist:1993ng}
  N.~A.~Tornqvist,
  From the deuteron to deusons, an analysis of deuteron - like meson meson bound states,
  Z.\ Phys.\ C {\bf 61}, 525 (1994)


\bibitem{Locher:1993cc}
  M.~P.~Locher, Y.~Lu and B.~S.~Zou,
  Rates for the reactions $\bar{p} p \to \pi \phi$ and $\gamma \phi$,
  Z.\ Phys.\ A {\bf 347}, 281 (1994)


\bibitem{Li:1996yn}
  X.~Q.~Li, D.~V.~Bugg and B.~S.~Zou,
 A Possible explanation of the '$\rho \pi$ puzzle' in $J/\psi$, $\psi^\prime$ decays,
  Phys.\ Rev.\ D {\bf 55}, 1421 (1997).

\bibitem{He:2019rva} 
  J.~He and D.~Y.~Chen,
  Molecular states from $\Sigma^{(*)}_c\bar{D}^{(*)}-\Lambda_c\bar{D}^{(*)}$ interaction,
  Eur.\ Phys.\ J.\ C {\bf 79}, no. 11, 887 (2019)



 \end{thebibliography}
\end{document}